\documentclass{aastex}

\usepackage{psfig,emulateapj5}

\newcommand{\noi}{\noindent}
\newcommand{\lsim}{\raisebox{0.3mm}{\em $\, <$} \hspace{-3.3mm}
\raisebox{-1.8mm}{\em $\sim \,$}}
\newcommand{\gsim}{\raisebox{0.3mm}{\em $\, >$} \hspace{-3.3mm}
\raisebox{-1.8mm}{\em $\sim \,$}}

\newcommand{\ex}{\rm e}

\shorttitle{Formation of Obscuring Walls by Starburst Radiation}
\shortauthors{Ohsuga \& Umemura}

\begin{document}


\title{Formation of Obscuring Walls by Radiation Force from Circumnuclear
Starbursts and Implications for Starburst-Active Galactic Nucleus
Connection}


\author{K. Ohsuga and M. Umemura}
\affil{Center for Computational Physics, University of Tsukuba,
Tsukuba, Ibaraki 305-8577, Japan}






\begin{abstract}
We explore the formation of dusty gas walls
induced by a circumnuclear starburst
around an active galactic nucleus (AGN). 
We concentrate our attention on the role of the radiation force by
a starburst as well as an AGN, where the effects of 
optical depth of dusty gas are taken into consideration.
First, we solve the hydrostatic equations in 
spherical symmetry coupled with the 
frequency-dependent radiative processes, to demonstrate 
that a geometrically thin, optically thick wall forms due to 
the radiation pressure by a circumnuclear starburst.
Next, in two-dimensional axisymmetric space,
we analyze the configuration and the stability of geometrically thin walls
which are in balance 
between radiation pressure and gravity.
%
As a result, it is shown that
the radiation force by the circumnuclear starburst works to stabilize 
optically thick walls surrounding the nucleus.
In the case of a brighter starburst with a fainter AGN
(e.g. $L_{\rm SB}/M_{\rm SB}\gsim 10[L_\odot/M_\odot]$
and $L_{\rm AGN}\lsim 10^{11} L_\odot$), 
there form double walls, an inner one of which
is located between the nucleus and the circumnuclear starburst, and
an outer one of which enshrouds
both the starburst regions and the nucleus.
The total extinction of both walls 
turns out to be larger for a brighter starburst,
which is $A_V\sim 10$ for $L_{\rm SB}/M_{\rm SB}\gsim 10^2[L_\odot/M_\odot]$.
As a consequence, double walls could heavily obscure the nucleus to make
the AGN type 2.
The outer wall may provide an explanation for 
the recent indications for large-scale obscuring materials in Seyfert 2's.
Also, it is predicted that the AGN type is time-dependent 
according to the stellar evolution in the starburst,
which shifts from type 2 to type 1 in several times $10^7$ yr
owing to the disappearance of walls.
In contrast, if the AGN itself is much brighter than the starburst
as a quasar is,
then neither wall forms regardless of the starburst activity 
and the nucleus is likely to be identified as type 1.
To conclude, the radiatively-supported gas walls could be responsible for 
the putative correlation between AGN type and the starbursts,
whereby Seyfert 2 galaxies are more frequently associated 
with circumnuclear starbursts than type 1, 
whereas quasars are mostly observed as type 1 regardless
of star-forming activity in the host galaxies.
\end{abstract}



\section{INTRODUCTION}

There has been a good deal of evidence on an obscuring torus of subparsec scale,
which surrounds an active galactic nuclei (AGN)
(Antonucci 1984; Wilson, Ward, \& Haniff 1988; Barthel 1989; 
Blanco, Ward, \& Wright 1990; Miller \& Goodrich 1990; 
Awaki et al. 1991; Storchi-Bergmann, Mulchaey, \& Wilson 1992).
The obscuring torus is thought to be responsible for the dichotomy
of AGN type in the context of the unified model (Antonucci 1993, for a review).
However, the origin and physical structure of the torus has not been well
elucidated, although some intriguing models are proposed
by Krolik \& Begelman (1988), Pier \& Krolik (1992a, 1992b, 1993),
Efstathiou \& Rowan-Robinson (1995),
and Manske, Henning, \& Men'shchikov (1998).

Recent observations on AGN hosts have gradually revealed that
the properties of host galaxies of Seyferts are 
intrinsically unlike between type 1 and type 2
(Heckman et al. 1989; Maiolino et al. 1995, 1997, 1998a;
P\'erez-Olea \& Colina 1996; Hunt et al. 1997; Malkan, Gorjian, \& Tam 1998; 
Storchi-Bergmann, Schmitt, \& Fernandes 1999;
Storchi-Bergmann et al. 2000;
Gonz\'alez Delgado, Heckman, \& Leitherer 2001).
The Seyfert 2 galaxies are more frequently associated with 
the circumnuclear starbursts than type 1's.
In contrast, quasars (QSOs) are mostly observed as type 1,
although the QSO hosts often exhibit vigorous star-formation activity
(Barvainis, Antonucci, \& Coleman 1992; Ohta et al. 1996; Omont et al. 1996;
Schinnerer, Eckart, \& Tacconi 1998; 
Brotherton et al. 1999; Canalizo \& Stockton 2000a, 2000b;
Dietrich \& Wilhelm-Erkens 2000).
[It is noted that the QSO hosts
are fainter than QSO nuclei themselves
(McLeod \& Rieke 1995b; Bahcall et al. 1997; 
Hooper, Impey, \& Foltz 1997; Crawford et al. 1999; Kirhakos et al. 1999;
McLure et al. 1999; McLure, Dunlop, \& Kukula 2000).]
%
Such correlation with circumnuclear starbursts 
seems beyond understanding based upon 
the picture of the unified model where
the bifurcation of AGN type is simply accounted for 
the orientation of the nucleus surrounded by an obscuring torus.

We have some significant pieces of information on the obscuring materials.
By X-ray observations, it is indicated
that most Seyfert 2 nuclei are heavily obscured along the line of sight
with at least $A_V>10$ and sometimes $A_V>100$ 
(Matt et al. 1996, 1999; Maiolino et al. 1998b; Bassani et al. 1999;
Risaliti, Maiolino, \& Salvati 1999).
On the other hand, $A_V$ of the nuclear or circumnuclear regions 
is estimated to be between a few and several
by IR and optical observations
(Rix et al. 1990; Roche et al. 1991; Goodrich, Veilleux, \& Hill 1994; 
McLeod \& Rieke 1995a; 
Oliva, Marconi, \& Moorwood 1999).
Also, it is argued that a component of obscuring materials
must be extended up to $\geq 100$ pc in addition to 
a compact component confined into subparsec scales
(Rudy, Cohen, \& Ake 1988; Miller, Goodrich, \& Mathews 1991; Goodrich 1995;
McLeod \& Rieke 1995a; 
Maiolino et al. 1995; Maiolino \& Rieke 1995; 
Malkan, Gorjian, \& Tam 1998).
These facts may suggest that the distributions of dusty gas around an AGN
are much more diverse than used to be considered, and 
they have a close relation with circumnuclear starburst events.

Recently, Ohsuga \& Umemura (1999, hereafter Paper I)
have suggested a novel picture for the starburst-AGN connection,
where a large-scale dusty wall of several hundred parsecs
is built up due to radiation force by a circumnuclear starburst 
as well as an AGN.
This model provides a possibility that an AGN type is regulated
by circumnuclear starbursts, and also it gives a physical origin
of extended obscuring materials.
However, the dusty wall was analyzed in an optically thin regime.
Thus, the possibility of the strong obscuration with $A_V \gg 1$ 
is still an open question in this picture. 


In this paper, by taking the optical depth of dusty gas into consideration,
we consider the effects of radiation force by a circumnuclear starburst
as well as an AGN, and investigate the stable configuration of dusty gas.
To begin with, we solve hydrostatic equations coupled with the
frequency-dependent radiative processes
with assuming spherical symmetry. 
By this analysis, we demonstrate that a geometrically thin, 
optically thick wall forms due to radiation
pressure by a starburst.
Next, we obtain the stable equilibrium configuration of dusty gas walls
in two-dimensional axisymmetric space.
Finally, we attempt to classify the AGN type according to 
the starburst luminosity and the AGN luminosity.
Moreover, taking account of the stellar evolution in starburst regions,
we discuss the time evolution of the circumnuclear structure.
In \S 2, the radiation fields and gravitational fields are modeled.
In \S 3, the structure of a dusty wall is analyzed in 
spherically symmetric approximation, demonstrating that
a geometrically thin, optically thick wall can form due to
radiation force by a starburst.
In \S 4,
the configuration of dusty gas in two-dimensional axisymmetric space
is investigated with an approximation of geometrically-thin walls,
and the conditions for the wall formation are given.
In \S 5, based upon the present picture,
we discuss the implications for the starburst-AGN connection.
Furthermore, the evolution of AGN type
which is predicted by the formation of the obscuring walls is shown in \S 6.
\S 7 is devoted to the conclusions.

\section{RADIATION FIELDS AND GRAVITATIONAL FIELDS}

Recent observations have been revealed that
the circumnuclear starburst regions frequently exhibit ring-like 
features 
(Wilson et al. 1991; Forbes et al. 1994; Marconi et al. 1994; 
Mauder et al. 1994; 
Buta, Purcell, \& Crocker 1995; Barth et al. 1995; Leitherer et al. 1996; 
Maoz et al. 1996; Storchi-Bergman, Wilson, \& Baldmin 1996).
Therefore, as a component of radiation sources,
we assume a starburst ring whose bolometric luminosity, 
radius, and total mass are $L_{\rm SB}$, $R_{\rm SB}$, and $M_{\rm SB}$,
respectively.
[Even if the starburst regions are localized, the radiation fields
could be equivalent to that by a ring-like source 
owing to the short rotation timescale (see also Paper I)].
We also include an AGN itself, in which the bolometric luminosity is $L_{\rm AGN}$
and we simply assume an energy spectrum of 
$L^{\rm AGN}_\nu \propto \nu^{-1}$ 
between 0.01 eV and 100 keV (Blandford, Netzer, \& Woltjer 1990).
To calculate the bolometric luminosity
and the energy spectrum of the starburst ring,
we assume the initial mass function (IMF) to be of Salpeter-type,
\begin{equation}
\phi \propto (m_*/M_{\odot})^{-1.35},
\end{equation}
and 
the star formation rate (SFR) in the starburst regions to be
\begin{equation}
{\rm SFR} \propto \exp(-\frac{t_{\rm SB}}{10^7 \rm yr}),
\end{equation}
where $t_{\rm SB}$ is the elapsed time after the initial starburst.
The total luminosity is regulated by
the mass-luminosity relation, $(l_*/L_{\odot})=(m_*/M_{\odot})^{3.7}$, 
and the mass-age relation, 
$\tau=1.1\times10^{10} {\rm yr}(m_*/M_{\odot})^{-2.7}$,
where $m_*$ and $l_*$ are respectively 
stellar mass and stellar luminosity.
Here, we consider the mass range of $2M_\odot-40M_\odot$ in the IMF,
since several observations indicate that the IMF is deficient in low mass
stars with the cutoff of about $2M_\odot$,
and the upper mass limit is inferred to be around $40M_{\odot}$
(Doyon, Puxley, \& Joseph 1992; Charlot et al. 1993;
Doane \& Mathews 1993; Hill et al. 1994; Brandl et al. 1996).
In addition, we employ the mass-temperature relation,
$(T_*/T_\odot)=(m_*/M_{\odot})^{0.575}$,
and the energy spectrum of the star is assumed to be
the blackbody spectrum with the effective temperature, $T_*$.
We neglect the radiation from the supernovae,
because the long term-averaged luminosity of supernovae is less than 10\%
of the total stellar luminosity of the ring as shown in Paper I.
Resultantly,
the bolometric luminosity of the starburst ring
is in proportion to the total mass,
and decreases with time.

As for the gravitational fields, we should consider the galactic bulge,
the central black hole, and the starburst ring.
But, the gravitational fields in circumnuclear regions
suffer from some uncertainty.
Recent observations provide significant information
which allows us to model the gravitational fields in circumnuclear regions.
For instance, the stellar rotation velocity in the Circinus galaxy
(a closest Seyfert 2 galaxy) 
shows that the mass distributions in circumnuclear regions 
are not concentrated into a point-like object but extended
to a few hundred parsecs (Maiolino et al. 1998a).
Therefore, we consider also such an inner bulge-like component.
We assume the galactic bulge to be an uniform sphere whose 
mass and radius are $M_{\rm GB}$ and $R_{\rm GB}$,
the inner bulge also to be an uniform sphere whose 
mass and radius are $M_{\rm IB}$ and $R_{\rm IB}$,
and the mass of the black hole to be $M_{\rm BH}$, respectively.

The observations of IRAS galaxies by Scoville et al. (1991) 
show that the central regions within kilo parsecs
possess the mass of $\lsim 10^{10} M_\odot$.
This is comparable to a typical mass of galactic bulge.
A rotation velocity of a starburst ring 
also implies that the dynamical mass within a few hundred parsecs is 
several $10^9 M_\odot$ (Elmouttie et al. 1998), which provides an estimation
of the mass of inner bulge.
In addition, Maiolino et al. (1998a) estimated an upper limit of a 
putative black hole as several $10^6M_\odot$.
(We tentatively assume the black hole of $10^7M_\odot$, but
the overall results are not changed much 
even if one assumes $M_{\rm BH}=10^6M_\odot$ or $10^8M_\odot$.)
By taking these observational data into account, we adopt the mass ratio 
as $M_{\rm GB}:M_{\rm IB}:M_{\rm BH}=1:0.1:10^{-3}$.
On the other hand, the mass of the starburst ring, $M_{\rm SB}$, 
is supposed to be between $10^7 M_\odot$ and $10^{10} M_\odot$.
Also, a starforming ring is observed around a few hundred 
parsecs in some galaxies 
(Marconi et al. 1994; Elmouttie et al. 1998).
Hence, here we assume a starburst ring at a radius of $200$ pc.
For the size ratios, we employ 
$R_{\rm GB}:R_{\rm SB}:R_{\rm IB}=5:1:0.5$.

\section{STRUCTURE OF RADIATIVELY-SUPPORTED OBSCURING WALL}

The equilibrium configuration of dusty gas is nearly
spherically symmetric as shown in the next section.
Thus, first we investigate the structure of a radiatively-supported
gas wall inside the starburst ring
under an approximation of local spherically symmetry.
For this purpose, we assume the gravity by the starburst ring to be 
constant at a part of the wall of interest.
Also, the radiation force by the starburst ring is assumed to to 
be a function solely of optical depth measured from 
the outer surface of the wall.
(Rigidly speaking, both assumptions are verified only when
the wall is geometrically thin and located at the equatorial plane.)
Under these assumptions,
we solve the one-dimensional radiation-hydrostatic equation,
coupled with ionization process and thermal process.

The radiation-hydrostatic equation is given by
\begin{eqnarray}
  -\frac{GM_l}{l^2}-\frac{1}{\rho_{\rm g}} \frac{dP_{\rm g}}{dl}
  + \int \frac{\chi_\nu}{c}
    \frac{L^{\rm AGN}_\nu \ex^{-\tau_\nu}}
  {4 \pi l^2} d\nu
  +f_{\rm grav}^{\rm SB} \nonumber\\
  -\int \frac{\chi_\nu}{c}
  F^{\rm SB}_\nu \ex^{-(\tau^{\rm total}_{\nu}-\tau_\nu)} d\nu=0,
\end{eqnarray}
where $l$ is the radius,
$M_l$ is the total mass within the radius $l$,
$\rho_{\rm g}$ is the gas density,
$P_{\rm g}$ is the gas pressure,
$\chi_\nu$ is the mass extinction coefficient of dusty gas,
$\tau_\nu$ is the optical depth of the wall measured from the center,
$\tau_\nu^{\rm total}$ is the total optical depth of the wall,
and $f_{\rm grav}^{\rm SB}$ and $F^{\rm SB}_\nu$ 
are respectively the gravity and the radiation flux by the starburst ring.
It is noted that $f_{\rm grav}^{\rm SB}$ and $F^{\rm SB}_\nu$
provide non-spherical components in real situations.
The effects of the non-sphericity are taken into consideration by
evaluating these quantities
by two-dimensional axisymmetric calculations.

Recent observations
report the existence of a large amount of dust,
molecular gas, and metals in QSOs
(Barvainis, Antonucci, \& Coleman 1992; Ohta et al. 1996; Omont et al. 1996;
Dietrich \& Wilhelm-Erkens 2000).
Therefore, we suppose the dust-to-gas mass ratio to be 0.03,
which is three times as large as that observed in the Solar neighborhood.
Since the mass density of dust and Thomson scattering are negligible,
the extinction is given by 
$\chi_\nu=(n_{\rm H} \chi_{\rm HI}\sigma^{\rm HI}_\nu+
\alpha_\nu^{\rm d})/\rho_{\rm g}$.
Here, $n_{\rm H}$ is the number density of hydrogen nuclei,
$\chi_{\rm HI}$ is the fraction of neutral hydrogen,
$\sigma^{\rm HI}_\nu$ is the photoionization cross-section,
and $\alpha_\nu^{\rm d}$ is the absorption coefficient of dust.
For the dust model, we employ the grain size distribution as
$n_{\rm d}(a_{\rm d}) \propto a_{\rm d}^{-3.5}$ in a range of
$[0.01\mu{\rm m}, 1\mu{\rm m}]$ (Mathis, Rumpl, \& Nordsieck 1977),
and absorption cross-section, 
$\pi a_{\rm d}^2 \min [1,(\lambda/2\pi a_{\rm d})^{-2}]$,
where $a_{\rm d}$ is the grain radius,
and the density of solid material within a grain is 
assumed to be $1.0 \rm g\, cm^{-3}$.

The equation of the ionization balance is
\begin{equation}
  \Gamma^\gamma \chi_{\rm HI} + \Gamma^{\rm ci} n_{\rm H}
  \chi_{\rm HI}(1-\chi_{\rm HI})
  =\alpha_{\rm rec} n_{\rm H} (1-\chi_{\rm HI})^2,
\end{equation}
where $\Gamma^\gamma$ is the photoionization rate,
$\Gamma^{\rm ci}$ is the collisional ionization rate,
and $\alpha_{\rm rec}$ is the recombination coefficient.
$\Gamma^\gamma$ is given by
\begin{equation}
  \Gamma^\gamma=
  \int_{\nu_{\rm L}} \sigma^{\rm HI}_\nu 
  \left\{
    \frac{L^{\rm AGN}_\nu \ex^{-\tau_\nu}}{4 \pi l^2}
    +F^{\rm SB}_\nu \ex^{-(\tau^{\rm total}_{\nu}-\tau_\nu)}
  \right\} \frac{1}{h\nu} d\nu,
\end{equation}
with $\nu_{\rm L}$ being the Lyman limit frequency,
and $\Gamma^{\rm ci}$ is 
$1.19\times 10^{-10}T_{\rm g}^{1/2}
\exp(-1.58\times 10^5/T_{\rm g})$ (Sherman 1979).
If a free electron recombines directly to ground state of hydrogen,
the emitted photon has enough energy to cause further photoionization.
Since a part of recombination photons are absorbed by the dust grain,
the value $\alpha_{\rm rec}$ is given by
$\alpha_{\rm rec} = \alpha_{\rm A}-(\alpha_{\rm A}-\alpha_{\rm B})
n_{\rm H} \chi_{\rm HI} \sigma^{\rm HI}_{\nu_{\rm L}}
/\rho_{\rm g}\chi_{\nu_{\rm L}}$,
where total recombination coefficient, $\alpha_{\rm A}$, is well fitted by
$\alpha_{\rm A}
=2.1\times 10^{-13}(T_{\rm g}/10^4 {\rm K})^{-1/2}
\phi(16T_{\rm g}/10^4 {\rm K})$,
with
$\phi(y)= 0.5(1.7+\ln y +1/6y)$ for $y\geq 0.5$, 
or $y(-0.3-1.2\ln y)+y^2(0.5-\ln y)$ for $y < 0.5$ (Sherman 1979),
and the recombination coefficient to all excited levels of hydrogen, 
$\alpha_{\rm B}$,
is approximately related with $\alpha_{\rm A}$ as
$\alpha_{\rm B}=\alpha_{\rm A} \exp \{ -0.487 
(T_{\rm g}/10^4 \rm K)^{1/5}\}$,
where $T_{\rm g}$ is the gas temperature. 

The equation for the energy balance of the gas is given by
\begin{equation}
  {\cal H}^{\gamma}_{\rm g}=
  {\cal L}^{\rm rec}_{\rm g}
  +{\cal L}^{\rm ff}_{\rm g}
  +{\cal L}^{\rm ci}_{\rm g}
  +{\cal L}^{\rm ce}_{\rm g}
  +{\cal L}^{\rm Z}_{\rm g}
  +{\cal L}^{\rm gd}_{\rm g},
\end{equation}
where ${\cal H}^{\gamma}_{\rm g}$
is the heating rate of the gas due to direct radiation by 
the starburst ring as well as the AGN, and
${\cal L}^{\rm rec}_{\rm g}$, ${\cal L}^{\rm ff}_{\rm g}$,
${\cal L}^{\rm ci}_{\rm g}$, ${\cal L}^{\rm ce}_{\rm g}$,
${\cal L}^{\rm Z}_{\rm g}$, and ${\cal L}^{\rm gd}_{\rm g}$, 
are respectively the cooling rates for gas through
radiative recombination, free-free emission, collisional ionization,
collisional excitation, emission for metal, and collision with dust.
${\cal H}^{\gamma}_{\rm g}$ is given by
\begin{eqnarray}
  {\cal H}^{\gamma}_{\rm g}=
  \int_{\nu_{\rm L}}   \chi_{\rm HI}  \sigma^{\rm HI}_\nu 
  \left\{
  \frac{L^{\rm AGN}_\nu \ex^{-\tau_\nu}}{4 \pi l^2} 
  +F^{\rm SB}_\nu \ex^{-(\tau^{\rm total}_{\nu}-\tau_\nu)}
  \right\} \nonumber\\
  \left(1-\frac{\nu_{\rm L}}{\nu}\right) d\nu,
\end{eqnarray}
where we neglect the energy transfer by the Compton scattering
because the timescale is estimated to be $\sim 10^9$ yr 
and it is much longer than the typical lifetime of AGN, 
$\sim 10^8$ yr, in the circumnuclear regions of $>$ several 10 pc
even in the case of very luminous AGNs with $\sim 10^{13}L_\odot$.
${\cal L}^{\rm rec}_{\rm g}$,
${\cal L}^{\rm ff}_{\rm g}$,
${\cal L}^{\rm ci}_{\rm g}$,
and ${\cal L}^{\rm ce}_{\rm g}$ are represented by
${\cal L}^{\rm rec}_{\rm g}
=3kT_{\rm g} \alpha_{\rm rec} n_{\rm H} (1-\chi_{\rm HI})^2/2 $,
${\cal L}^{\rm ff}_{\rm g}=1.42\times 10^{-27}T_{\rm g}^{1/2}
n_{\rm H} (1-\chi_{\rm HI})^2$,
${\cal L}^{\rm ci}_{\rm g}=\Gamma^{\rm ci} h\nu_{\rm L}
n_{\rm H} \chi_{\rm HI}(1-\chi_{\rm HI})$,
${\cal L}^{\rm ce}_{\rm g}=7.5\times 10^{-19}
\exp( -1.18\times 10^5/T_{\rm g})
n_{\rm H} \chi_{\rm HI}(1-\chi_{\rm HI})$ (Black 1981),
and ${\cal L}^{\rm Z}_{\rm g}=4\times 10^{-28}T_{\rm g}^{1/2}
n_{\rm H}$ for $10^2{\rm K}<T_{\rm g}<10^4{\rm K}$, 
or ${\cal L}^{\rm Z}_{\rm g}=10^{-22.1-120Z} T_{\rm g}^{30Z}
n_{\rm H}-10^{-22.1}n_{\rm H}$
for $10^4{\rm K}<T_{\rm g}<10^5{\rm K}$ (Theis, Burkert, \& Hensler 1992),
where $Z$ is the metal abundance.
Here, the metal abundance is assumed to be $3Z_\odot$,
consistently with the supposed dust-to-gas mass ratio, 0.03.
Assuming the collisional equilibrium of charge on a grain,
we can estimate
${\cal L}_{\rm g}^{\rm gd}  
= \int n_{\rm d} \pi a_{\rm d}^2 da_{\rm d} \cdot
(7-6\chi_{\rm HI})(8kT_{\rm g}/\pi m_{\rm p})^{1/2}
(2kT_{\rm g}-2kT_{\rm d})$,
where $m_{\rm p}$ is the proton mass and
$T_{\rm d}$ is the dust temperature.
As for the energy balance for dust, the emission from gas 
is assumed to be absorbed by dust almost on the spot.
Then, the energy equation for dust is
\begin{equation}
  {\cal H}^{\gamma}_{\rm d}
  +{\acute {\cal L}}^{\rm rec}_{\rm g}
  +{\cal L}^{\rm ff}_{\rm g}
  +{\cal L}^{\rm ce}_{\rm g}
  +{\cal L}^{\rm Z}_{\rm g}
  +{\cal L}^{\rm gd}_{\rm g}
  ={\cal L}^{\gamma}_{\rm d},
\end{equation}
where ${\cal H}^{\gamma}_{\rm d}$ and ${\acute {\cal L}}^{\rm rec}_{\rm g}$ 
are the heating rates for dust 
due to the direct radiation from a starburst ring as well as an AGN 
and due to the recombination photons from hydrogen,
and ${\cal L}^{\gamma}_{\rm d}$ is the cooling by dust emission.
(This assumption is justified by the fact that a resultant gas wall
is optically thick due to dust opacity.)
${\cal H}^{\gamma}_{\rm d}$ is given by
\begin{equation}
  {\cal H}^{\gamma}_{\rm d}= \frac{1}{n_{\rm H}}
  \int \alpha_\nu^{\rm d}
  \left\{
  \frac{L^{\rm AGN}_\nu \ex^{-\tau_\nu}}{4 \pi l^2} 
  +F^{\rm SB}_\nu \ex^{-(\tau^{\rm total}_{\nu}-\tau_\nu)}
  \right\} d\nu,
\end{equation}
${\acute {\cal L}}^{\rm rec}_{\rm g}$ is written as
${\acute {\cal L}}^{\rm rec}_{\rm g}
=(3kT_{\rm g}/2 +h\nu_{\rm L})\alpha_{\rm rec} n_{\rm H} (1-\chi_{\rm HI})^2 $,
and ${\cal L}^{\gamma}_{\rm d}$ is presented by
${\cal L}^{\gamma}_{\rm d}= 
(\sigma \alpha^{\rm d}_{\nu_0}/\pi n_{\rm H})$
$\{ \Gamma (6) \zeta (6)/\Gamma (4) \zeta (4)\}
(k/h\nu_0)^2  T_{\rm d}^{6}$,
where $\Gamma$ and $\zeta$ are respectively the gamma function 
and Riemann's function (Nakamoto \& Nakagawa 1994).
We note that dust is primarily heated by the direct radiation, 
so that the dust temperature is not affected heavily
even if the heating by the gas emission is dismissed.
Moreover, since the absorption cross-section against IR radiation
re-emitted from dust is much smaller
than that against optical or UV radiation,
we suppose that dust grains work as pure absorbers.

The above equations and the equation of state for ideal gas are solved
to give the gas density, the gas temperature, the fraction of neutral hydrogen,
and the dust temperature as functions of radii.
The resultant profile of hydrogen number density and 
$A_V$ measured from the center are shown in Figure 1.
Here, $L_{\rm AGN}=10^{10}L_\odot$, $M_{\rm SB}=10^9M_\odot$,
and $L_{\rm SB}=8.3\times 10^{10}L_\odot$ $(t_{\rm SB}=3\times 10^7$ yr)
are assumed.
As seen in this figure, there forms an optically-thick wall, where
the inner surface is pushed outward by the radiation force by the AGN
against the gravity by the inner 
\centerline{\psfig{file=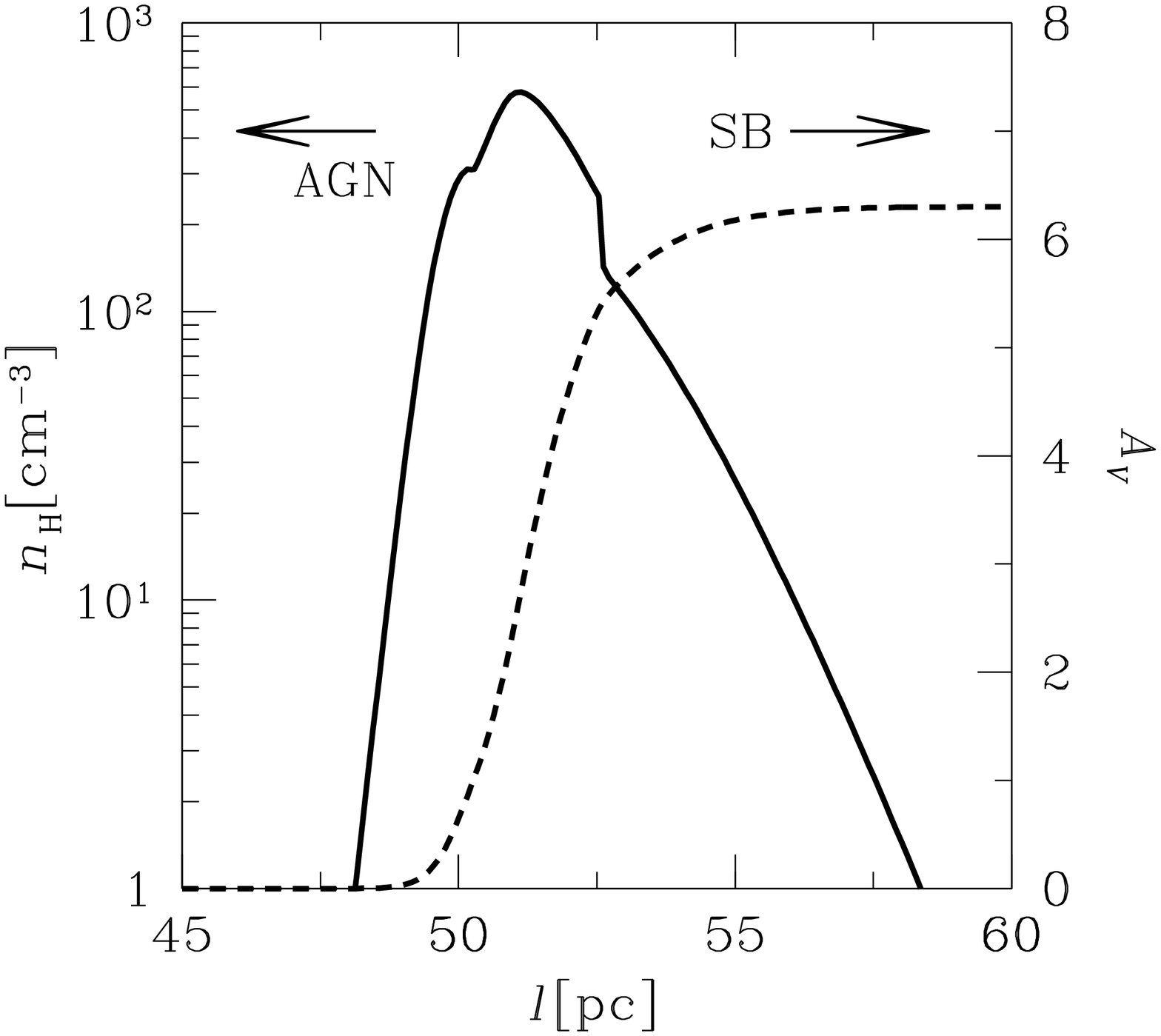,width=3.4in}}
\figcaption[fig1.eps] {
The profile of hydrogen number density
and the visual extinction $A_V$ measured from the center 
are shown as functions of radii, $l$.
Here, $L_{\rm AGN}=10^{10}L_\odot$, $M_{\rm SB}=10^9M_\odot$,
and $L_{\rm SB}=8.3\times 10^{10}L_\odot$ ($t_{\rm SB}=3\times 10^7$ yr)
are assumed.
The AGN is located at the center, $l=0$, and 
the size of the starburst ring is assumed to be 200 pc.
The interior and exterior surface of the wall are irradiated by
the radiation from the AGN and the starburst ring, respectively.
The density gradient is very steep
since the wall is compressed due to strong radiation force by
the starburst ring as well as the AGN.
The effective thickness of the wall is less than 10\% of its radial extension.
This figure shows that a geometrically thin, optically thick wall forms.
\label{fig1}}
\noi
bulge
and the radiation force by the starburst shoves the wall inward.
The optical depth ($A_V$) is basically determined by the ratio of 
the net radiation force to the total gravity if both AGN and
starburst are super-Eddington respectively to $M_l$ and to $M_{\rm SB}$.
The density profile of the wall has a peak at $\sim 50$ pc
and the density gradients on both sides of a peak are very steep.
Then, the geometrical thickness of the wall is basically determined 
by the ratio of the square of sound speed to the total radiation force.
The gas temperature is around $10^4$K due to the photoheating.
As a result, the effective thickness of the wall is less than 10\% of 
its radial extension. Hence, it is concluded that
a radiatively-supported wall can be optically thick and geometrically thin.
If the starburst is more luminous, the wall would be thinner
because the density gradient would be steeper
owing to the radiation force.
On the other hand, if the mass of the starburst ring is less than 
$10^9M_\odot$, the density profile would be roughly the same,
since the force fields are mainly determined by the AGN and the inner bulge.
In this analysis it has turned out 
that the radiation force from the AGN as well as the starburst
is exerted mainly on dust. 
Therefore, for the mass extinction of the dusty gas,
we take only the dust opacity into account in two-dimensional calculations 
below.

Although we have considered the structure of a gas wall inside 
the starburst ring, the structure of 
an outer wall beyond the starburst ring can be
understood by a similar argument. If an outer wall forms at several
hundred parsecs, the photoheating does not work effectively
to raise the temperature $\sim 10^4$K, but it
becomes around $\sim 10^2$K. Then, the thickness of the wall is
much thinner than the inner wall. As for the optical depth ($A_V$),
it is determined basically by the ratio of the radiation force to
the total gravity for super-Eddington luminosity, and
therefore could be $A_V \gg 1$.

\section{CONFIGURATION OF OBSCURING WALLS}

In this section, we examine the equilibrium configuration of a gas wall 
and the stability in two-dimensional axisymmetric space.
We assume the wall to be geometrically thin as analyzed 
in the previous section. Under the assumption, 
taking the effects by the optical depth into account,
we calculate the radiation force and the gravity
which are exerted on a dusty wall.
The radiative flux force by the starburst ring is given by
\begin{equation}
  f^i_{\rm SB} = \int \frac{\bar{\chi}_{\rm SB}}{c}
  \frac{\rho_{\rm SB}}{4\pi l_{\rm SB}^2}
  \frac{1-\exp(-\tau_{\rm SB}/\cos\theta_{\rm SB})}
  {\tau_{\rm SB}/\cos\theta_{\rm SB}} n^idV,
  \label{force_SB}
\end{equation}
at a point of $(r, z)$ in cylindrical coordinates,
where $i$ denotes $r$ or $z$,
$\bar{\chi}_{\rm SB}$ is the flux mean mass extinction coefficient
averaged over the starburst radiation,
$\tau_{\rm SB}$ is the total optical depth of the wall
which is estimated with using $\bar{\chi}_{\rm SB}$,
$\rho_{\rm SB}$ is the luminosity density in the starburst ring,
$l_{\rm SB}$ is the distance from $(r,z)$ to 
a volume element $dV$ of the ring,
$\theta_{\rm SB}$ is the viewing angle from this element,
and $n^i$ is the directional cosine.
Similarly, the radiative flux force by the AGN is presented by
\begin{equation}
  f^i_{\rm AGN} = \frac{\bar{\chi}_{\rm AGN}}{c}
  \frac{L_{\rm AGN}}{4\pi l^2}
  \frac{1-\exp(-\tau_{\rm AGN}/\cos\theta_{\rm AGN})}
  {\tau_{\rm AGN}/\cos\theta_{\rm AGN}} \frac{i}{l},
  \label{force_AGN}
\end{equation}
at the same point,
where $\bar{\chi}_{\rm AGN}$ is the flux mean mass extinction
averaged over the AGN radiation, $l$ is the distance, 
$\theta_{\rm AGN}$ is the viewing angle from the center, and
$\tau_{\rm AGN}$ is the total optical depth,
which is estimated with using $\bar{\chi}_{\rm AGN}$.
Using equations (\ref{force_SB}) and (\ref{force_AGN}),
the equilibrium between the radiation force and the gravity
is determined by
\begin{equation}
  -\frac{d \Phi}{dz}=f^z_{\rm SB}+f^z_{\rm AGN}+f^z_{\rm grav}=0,
  \label{zbalance}
\end{equation}
in the vertical directions ($z$-directions) and 
\begin{equation}
  -\frac{d \Phi}{dr}=
  \frac{j^2}{r^3}+f^r_{\rm SB}+f^r_{\rm AGN}+f^r_{\rm grav}=0,
  \label{rbalance}
\end{equation}
in the radial directions ($r$-directions), where 
$\Phi$ is the effective potential,
$j$ is the specific angular momentum of the dusty gas and
$f^i_{\rm grav}$ is the gravitational force.
Furthermore, stable configuration should satisfy the conditions
\begin{equation}
  \frac{d^2 \Phi}{dz^2} > 0,
  \label{zstable}
\end{equation}
and 
\begin{equation}
  \frac{d^2 \Phi}{dr^2} > 0.
  \label{rstable}
\end{equation}


\subsection{Inner Obscuring Wall}

First, we consider the formation of a dusty wall inside the starburst ring.
The wall is irradiated outward by the AGN and inward by the starburst ring.
In Figure 2, the resultant equilibrium branches
are shown in the $r$-$z$ space for $A_V=3$, $5$, and $7$.
Here, $L_{\rm AGN}=10^{10}L_\odot$, $M_{\rm SB}=10^9M_\odot$,
and  $L_{\rm SB}=8.3\times 10^{10}L_\odot$ $(t_{\rm SB}=3\times 10^7$ yr)
are assumed.
On each curve,
the conditions for stable equilibrium in the vertical directions
[(\ref{zbalance}) and (\ref{zstable})] are satisfied.
However, the dashed curves do not satisfy the force balance (\ref{rbalance})
in the radial directions,
so that the dusty gas is likely to be swung away around the curves because of 
centrifugal force.
%
%
As a result, it is found that the solid curves are 
the stable equilibrium branches in both vertical and radial directions,
which provide the final configuration of the inner obscuring wall.
%
%
It is noted that the starburst ring plays an important role
for the force balance and the stability of the wall.
%
%
The AGN as well as the bulge components provides only spherically symmetric
forces, whereas the radiation force and the gravity by the starburst ring
are not spherically symmetric.
Therefore, on the tangential plane, the azimuthal component of 
centrifugal force can balance only with that of 
radiation force by the ring (see Figure 3).
To realize this situation, the starburst luminosity is required to 
be super-Eddington, $L_{\rm SB}>4 \pi cGM_{\rm SB}/\bar\chi_{\rm SB}$.
Since the effective radiation force by the starburst ring is weakened
for a large viewing angle $\theta_{\rm SB}$,
the radial balance breaks down in the vicinity of the
rotation axis (z-axis).
\centerline{\psfig{file=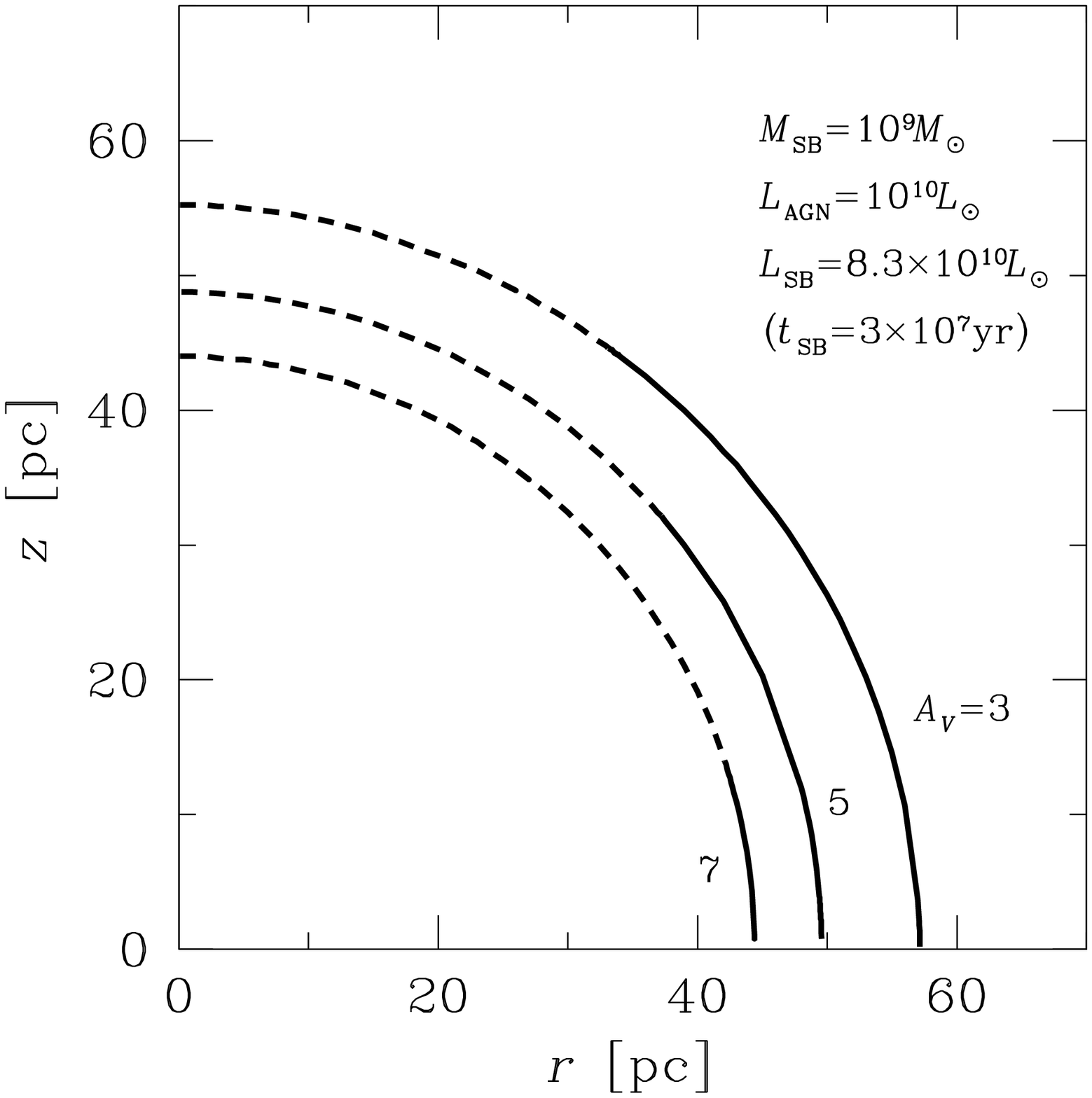,width=3.4in}}
\figcaption[fig2.eps] {
Equilibrium configuration of inner walls 
is shown in $r$-$z$ space for $A_V=3$, $5$, and $7$.
Here, $L_{\rm AGN}=10^{10}L_\odot$, $M_{\rm SB}=10^9M_\odot$,
and $L_{\rm SB}=8.3\times 10^{10}L_\odot$ ($t_{\rm SB}=3\times 10^7$ yr)
are assumed.
The solid curves represent the stable branches,
while the dashed curves are radially nonequilibrium branches.
The solid curves show the final configuration of 
stable inner walls.
Since the effective radiation force by the ring is weakened
in the vicinity of the rotation axis in the case of the inner wall
due to a large viewing angle,
an opening forms and it is shown by a dashed curve.
\label{fig2}}
\medskip
\centerline{\psfig{file=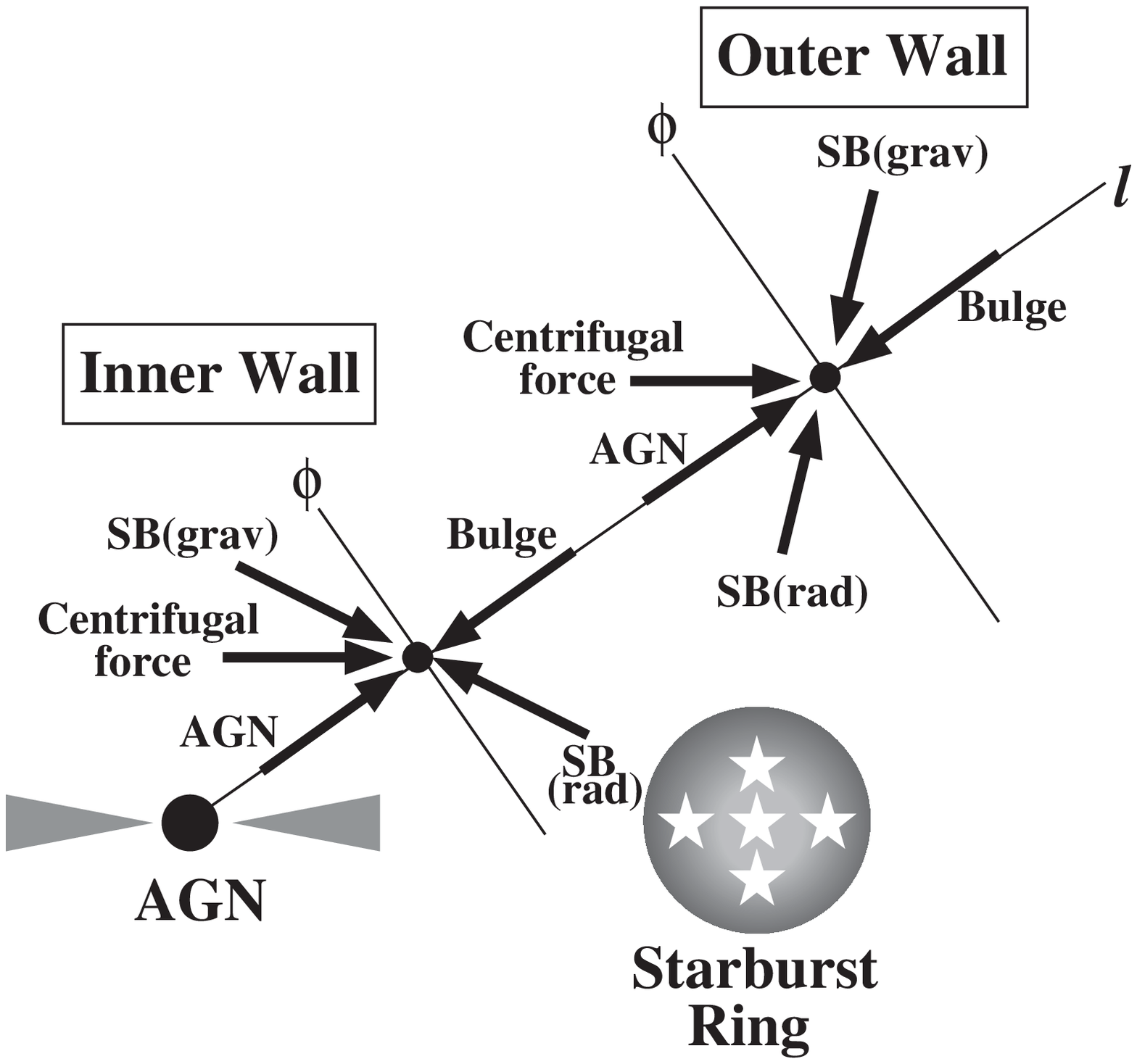,width=3.4in}}
\figcaption[fig3.eps] {
Schematic edge-on sketch showing the forces exerted on the inner wall
as well as the outer wall.
Arrows represent the orientations of forces,
although the lengths of the arrows do not correspond to the strength of forces.
As a result of numerical calculation,
it is found that the starburst ring provides nonspherical radiation force
and gravity,
whereas the forces by the AGN as well as the bulge components 
are spherically symmetric.
Hence, the walls form if the azimuthal ($\phi$) component of radiation force 
by the starburst ring balances with that of total force of 
the gravity by the ring and the centrifugal force.
\label{fig3}}

\medskip
\noi
An opening grows as the optical depth ($A_V$) increases,
because of the dilution of the effective radiation force by the ring
[see equation (\ref{force_SB})].
In order that a stable inner wall enshrouds the nuclear regions
with a large covering factor 
(opening angle $< 90^\circ$), 
$A_V$ is constrained from above as
\begin{equation}
  A_V \lsim \frac{\chi_V L_{\rm SB}}{4 \pi cGM_{\rm SB}},
  \label{smallAV}
\end{equation}
where $\chi_V$ is the mass extinction at the $V$ band.
On the other hand, the radiation force by a very luminous AGN could blow
the dusty gas away from the inner bulge regions
if the optical depth is not enough to dilute the radiation force.
This provides a lower bound of $A_V$ as
\begin{equation}
   A_V > \frac{\chi_V L_{\rm AGN}}{4 \pi cG
    \left(M_{\rm IB}+M_{\rm BH}\right)}.
  \label{smallAV2}
\end{equation}
Because of the time dependence of starburst luminosity,
the upper bound of $A_V$ given by inequality (\ref{smallAV})
is a decreasing function of time.
For a given model, the allowed range of $A_V$ of the inner wall
is presented in the Figure 4 as a function of time, $t_{\rm SB}$.
In the shaded regions, 
no 
\centerline{\psfig{file=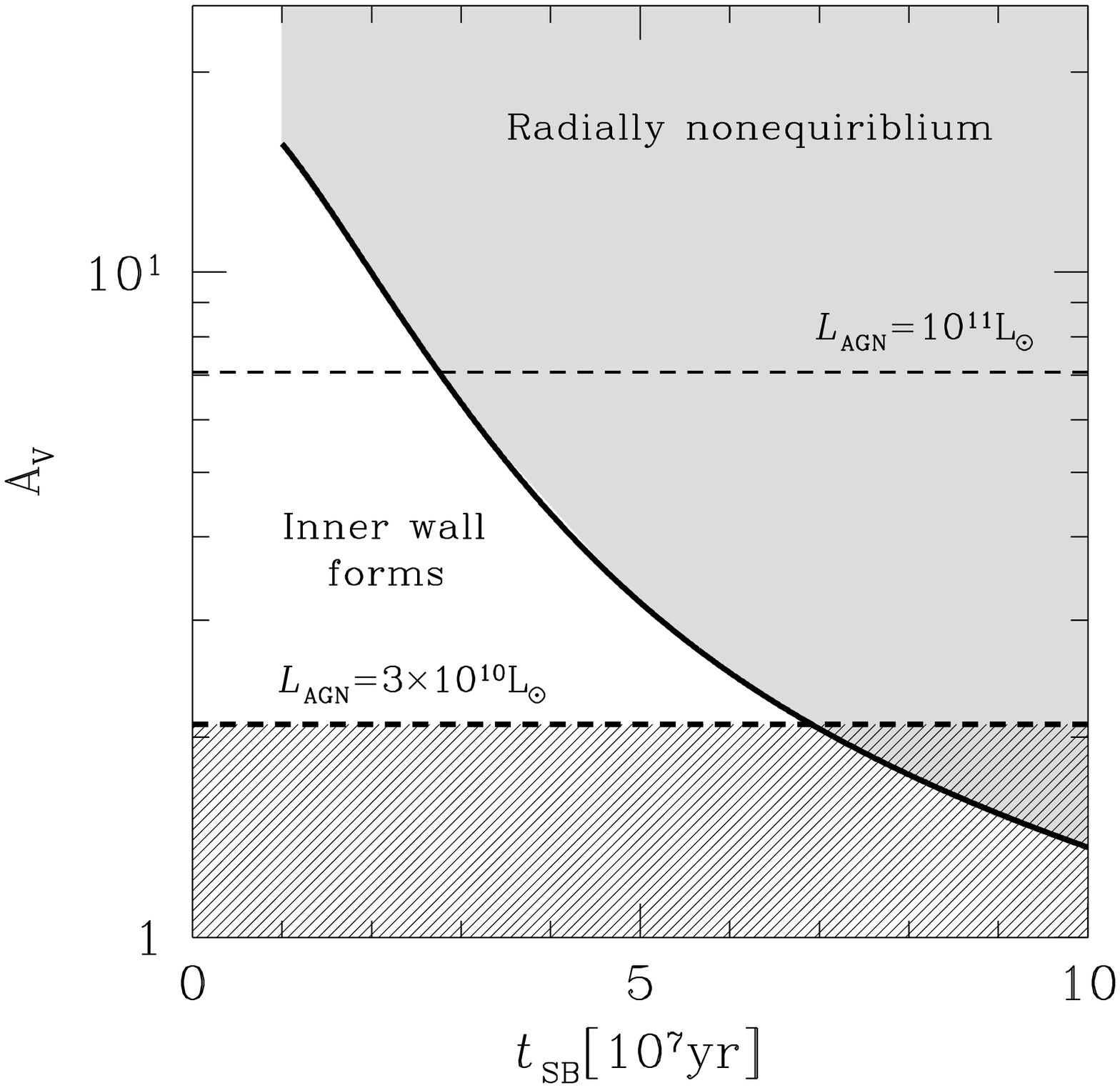,width=3.4in}}
\figcaption[fig4.eps] {
The range of $A_V$ allowed for the inner wall formation.
In the shaded area, $A_V$ does not satisfy the inequality (\ref{smallAV}),
so that the greater part of the wall is out of radial equilibrium. 
Also, the dusty gas is blown away due to strong radiation force by
the AGN in the hatched area in the case of 
$L_{\rm AGN}=3\times 10^{10}L_\odot$.
\label{fig4}}
\medskip
\noi
stable equilibrium branches with a large covering factor 
exist and in the hatched regions
the radiation force by the AGN blows out the dusty gas.
If the upper bound in (\ref{smallAV}) is smaller 
than the lower bound in (\ref{smallAV2}),
there is no allowed value for $A_V$.
That is to say, 
no inner wall forms regardless of $t_{\rm SB}$ if
\begin{equation}
  L_{\rm AGN}> \left( M_{\rm IB}+M_{\rm BH} \right)
  \frac{L_{\rm SB}}{M_{\rm SB}}.
 \label{consmall}
\end{equation}

\subsection{Outer Obscuring Wall}

Next, we investigate the formation of an outer wall with the radial
extension of $\geq R_{\rm SB}$.
In Figure 5, the resultant equilibrium branches of the outer wall
are shown in the $r$-$z$ plane.
As shown in this figure, a nearly spherical wall forms.
In this case,
the starburst ring as well as the AGN irradiate the inside of the wall.
Therefore, the radiation force by the starburst ring is exerted outward 
\centerline{\psfig{file=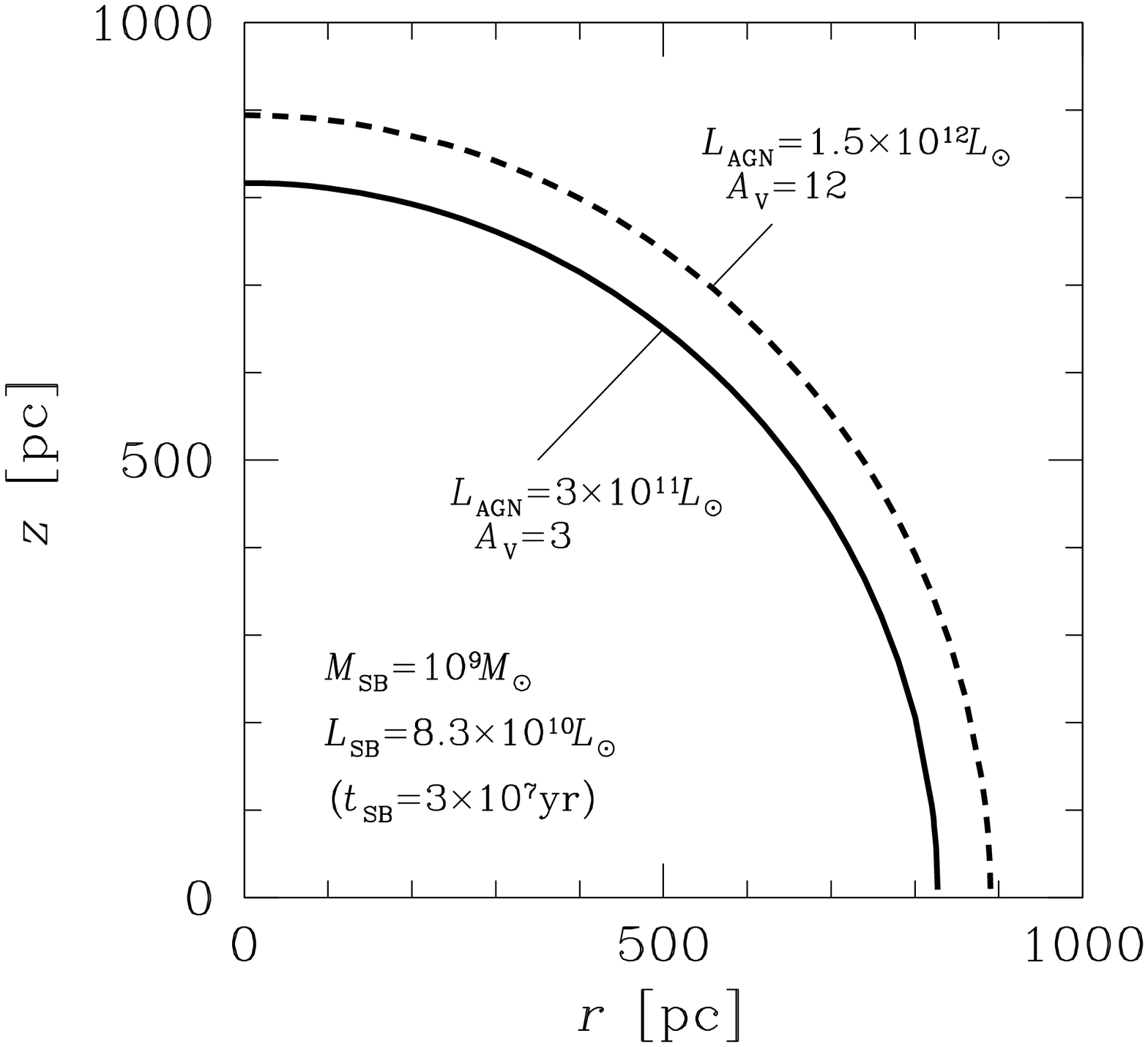,width=3.4in}}
\figcaption[fig5.eps] {
Equilibrium configuration of an outer wall 
beyond the starburst regions is shown in $r$-$z$ space
for two cases of the AGN luminosity, say, $L_{\rm AGN}=3\times 10^{11}L_\odot$
and $1.5\times 10^{12}L_\odot$.
Here, $M_{\rm SB}=10^9M_\odot$ and $L_{\rm SB}=8.3\times 10^{10}L_\odot$
are assumed.
The solid curves represent stable branches,
while the dashed curves are radially nonequilibrium branches.
The outer wall is roughly of spherical shape
and covers a wide solid angle whenever it forms.
\label{fig5}}
\medskip
\noi
on the dusty wall,
although the inner wall is pushed inward by the starburst radiation.
But, it has been found 
by the present analyses that
the radial equilibrium and stable condition for the outer wall formation is 
again given by the inequality (\ref{smallAV}) 
for a wide range of parameters.
It is also required that the starburst ring is super-Eddington.
This can be also understood by the argument of force balance in
the azimuthal directions at a point on the wall. 
The azimuthal component of the centrifugal force
can be balanced only by the force by
the starburst ring, because only the starburst ring provides
a spherical force (see Figure 3).  
The upper value of the allowed $A_V$ decreases again with time
as the starburst becomes dimmer.
%
As a prerequisite condition for the formation of the wall,
$(\bar\chi_{\rm AGN}L_{\rm AGN}+\bar\chi_{\rm SB}L_{\rm SB})
/4 \pi cG(M_{\rm SB}+M_{\rm IB}+M_{\rm BH})>1$ is required,
otherwise the dusty gas is not held at the regions of $> R_{\rm IB}$
even in the case of an optically-thin wall.
Since the radiation force is diluted by the optical depth of the wall,
the upper bound of $A_V$ is provided as
\begin{equation}
  A_V <
  \frac{ \chi_V (L_{\rm AGN}+L_{\rm SB})}{4 \pi cG
    (M_{\rm SB}+M_{\rm IB}+M_{\rm BH})},
\end{equation}
for the sustainment of the dusty gas.
On the other hand,
the lower bound of $A_V$ is estimated by 
\begin{equation}
  A_V > \frac{ \chi_V (L_{\rm AGN}+L_{\rm SB})}{4 \pi cG
    (M_{\rm GB}+M_{\rm SB}+M_{\rm IB}+M_{\rm BH})},
\end{equation}
since the radiation force would blow out the dusty gas in most regions
if the $A_V$ of the wall is smaller than this value.
Comparing this lower limit to equation (\ref{smallAV}), 
we find that no outer wall 
forms if the AGN luminosity satisfies the inequality as
\begin{equation}
  L_{\rm AGN}>(M_{\rm GB}+M_{\rm IB}+M_{\rm BH})\frac{L_{\rm SB}}{M_{\rm SB}}.
  \label{conlarge}
\end{equation}
It is noted that this condition involves the prohibition condition 
(\ref{consmall}) for an inner wall. 
$A_V$ is limited from above as
\begin{equation}
  A_V < \min \left[
    \frac{\chi_V L_{\rm SB}}{4 \pi cGM_{\rm SB}},
    \frac{\chi_V \left( L_{\rm AGN}+L_{\rm SB} \right)}
    {4 \pi cG(M_{\rm SB}+M_{\rm IB}+M_{\rm BH})} \right].
  \label{minout}
\end{equation}

\section{IMPLICATIONS FOR STARBURST-AGN CONNECTION}

From a viewpoint of the formation of radiatively-supported obscuring walls,
we discuss the connection between the AGN type and the starburst events.
Based upon the conditions of (\ref{consmall}) and (\ref{conlarge}),
in Figure 6 the prediction for the wall formation is summarized
in a diagram of $L_{\rm AGN}$ versus $L_{\rm SB}/M_{\rm SB, 9}$,
where $M_{\rm SB, 9}$ is the mass of a starburst in units of 
$10^9M_\odot$.
An outer wall forms below a solid line 
and 
an inner wall as well as an outer wall forms below a dashed line.
A vertical dot-dashed line shows the boundary for the formation
of the walls. The possible values of $A_V$ are also shown by
thin dot-dashed lines.
Here, it is stressed that 
the possible $A_V$ is practically determined by equation (\ref{minout})
for the outer wall.
However, 
if the mass supply to an altitude of several 100 pc 
by a superbubble driven by a 
circumnuclear starburst is insufficient,
no outer wall of $A_V \gsim$ several might form
(see next section).
The figure shows that for a luminous starburst and a relatively 
faint AGN,
both inner and outer walls are built up, and 
\centerline{\psfig{file=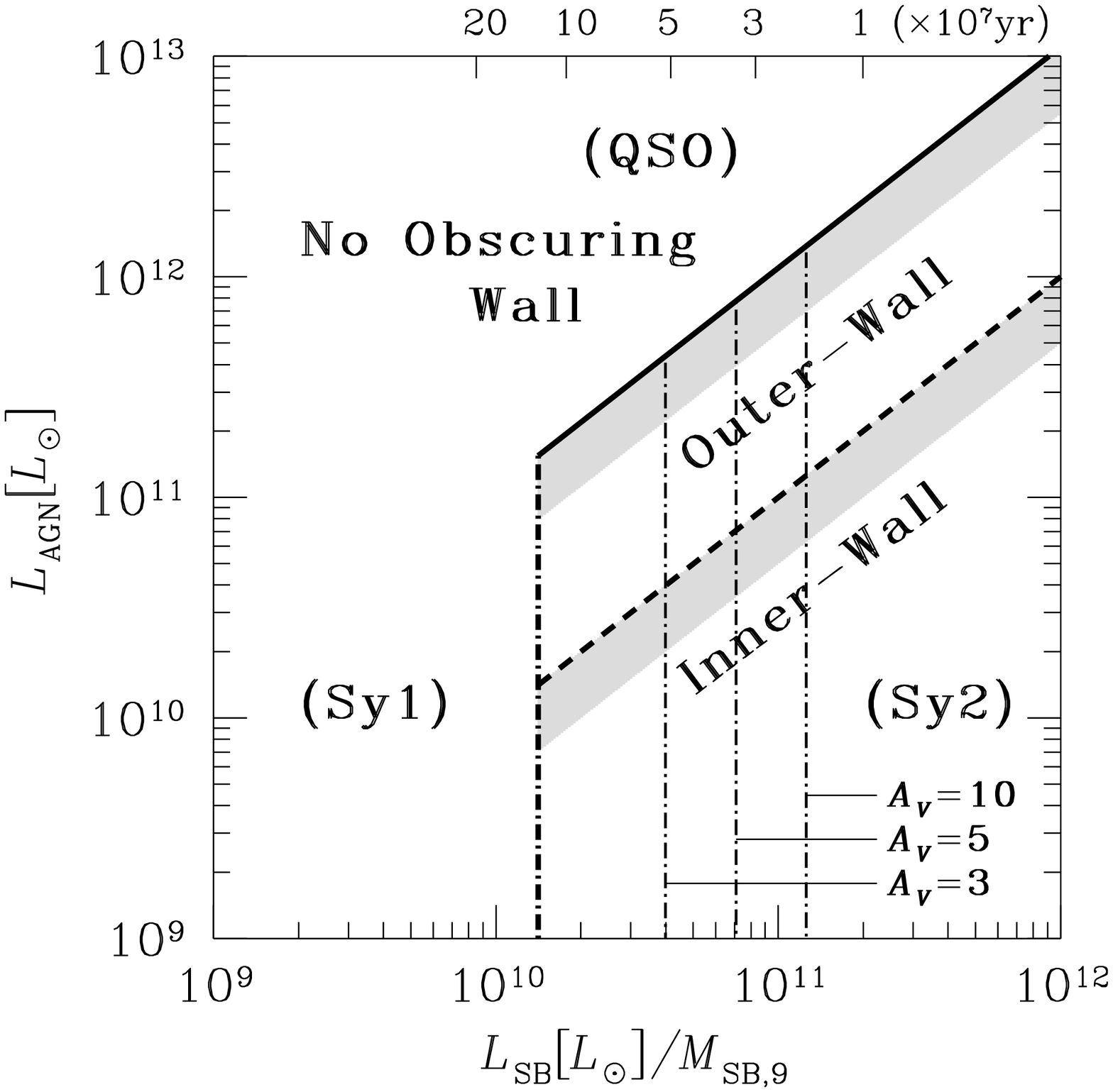,width=3.4in}}
\figcaption[fig6.eps] {
The conditions for the formation of obscuring walls are shown
in a diagram of the AGN luminosity $L_{\rm AGN}$ versus the starburst
luminosity $L_{\rm SB}$, where the starburst
luminosity is divided by the mass of starburst in units of $10^9M_\odot$.
An outer wall forms below a solid line and
an inner wall as well as an outer wall forms below a dashed line.
A vertical dot-dashed line gives the boundary for the formation
of the walls based on the condition that the starburst is super-Eddington,
and the specific values of $A_V$ are also shown by
thin dot-dashed lines.
The figure shows that for a luminous starburst and a relatively faint AGN,
the nucleus is enshrouded by double walls with the total extinction
of $A_V>10$, and therefore it is likely to be identified as a type 2 AGN (Sy 2).
If a starburst is intrinsically faint or becomes dimmer owing to
the stellar evolution, neither wall forms and the nucleus 
is observed as a type 1 AGN (Sy 1).
The age of the starburst based on the present stellar evolution model
is shown in the upper abscissa.
Furthermore, if the AGN is much more luminous, 
the wall formation is prohibited regardless of the
starburst luminosity (QSO).
\label{fig6}}
\noi
the nucleus
could be obscured with the total extinction
of $A_V\sim 10$. 
Then, the nucleus is likely to be identified as a type 2 AGN (Sy 2).
Also, the outer wall may be consistent with the recently observed
obscuring material extending up to $\geq 100$ pc around the nuclei
(Rudy, Cohen, \& Ake 1988; Miller, Goodrich, \& Mathews 1991; Goodrich 1995;
McLeod \& Rieke 1995a; 
Maiolino et al. 1995; Maiolino \& Rieke 1995; 
Malkan, Gorjian, \& Tam 1998).

By X-ray observations, 
the $A_V$ of most Seyfert 2 nucleus is estimated to be
$> 10$ and sometimes $> 100$.
Such a difference
between $A_V$ estimated by the X-ray observations
and $A_V$ assessed by the IR or Optical observations might be 
due to obscuring matter free from dust,
located just inside the dust sublimation radius of subparsec.
The structure and formation mechanism of obscuring material in 
subparsec regions is debatable, 
but it is beyond the scope of this paper.

If a starburst is intrinsically faint,
neither an outer wall nor inner one 
is expected to form. Then, the nucleus could be observed as a type
1 AGN (Sy 1). These results provide a physical explanation
for the putative correlation between AGN type and host properties whereby
Seyfert 2 galaxies 
are more frequently associated with circumnuclear starbursts
than type 1 galaxies. In the cases of the moderate extinction by the walls
may lead to intermediate types of AGNs, that is, type 1.9, 1.8,
1.5, or 1.2 according to the decrease of the extinction.

In addition, we have come to a significant conclusion that
a much higher AGN activity as (\ref{conlarge}) 
would preclude either wall from forming.
Then the nucleus
tends to be identified as type 1 regardless of the 
starburst luminosity. 
This result may be closely related to the fact that
QSOs are mostly observed as type 1 regardless
of star-forming activity in the host galaxies.

\section{AGN EVOLUTION}
In the previous section, it is shown that 
the obscuring walls form in the case of the luminous starburst and
the relatively faint AGN.
It implies that 
the obscuration of the AGN is time-dependent,
since the bolometric luminosity of the starburst ring decreases with time 
due to stellar evolution.
In this section,
we discuss the evolution of the AGN type predicted by this picture.
The schematic view of the evolution of circumnuclear structure 
is shown in Figure 7.
%

In an early evolutionary stage,
a large amount of dusty gas could be blown out by superbubbles
in the circumnuclear starburst regions.
Shapiro \& Field (1976),
Tomisaka \& Ikeuchi (1986), and Norman \& Ikeuchi (1989) have
shown that a superbubble of $\sim$ 1kpc
powered by massive OB associations forms.
As an observational example,
the superbubble of 1.1kpc is observed 
in the immediately east of the nucleus of 
nearby Seyfert 2/LINER galaxy NGC3079 (Veilleux et al. 1994).

Also, 
if the mass supply rate given by Norman \& Ikeuchi (1989) is applied to 
the case of circumnuclear starbursts of $M_{\rm SB} \sim 10^9 M_\odot$,
the rate into the regions of several 100 pc height
from the circumnuclear molecular disk due to superbubbles is estimated as
several $\times 10 M_\odot \,{\rm yr}^{-1}$.
In the case of the superbubble of the NGC3079,
Duric \& Seaquist (1988) and Veilleux et al. (1994) 
evaluated the mass supply rate to be $\gsim 10M_\odot\,{\rm yr}^{-1}$ 
and $\sim 7M_\odot\,{\rm yr}^{-1}$, respectively.
As a 
\centerline{\psfig{file=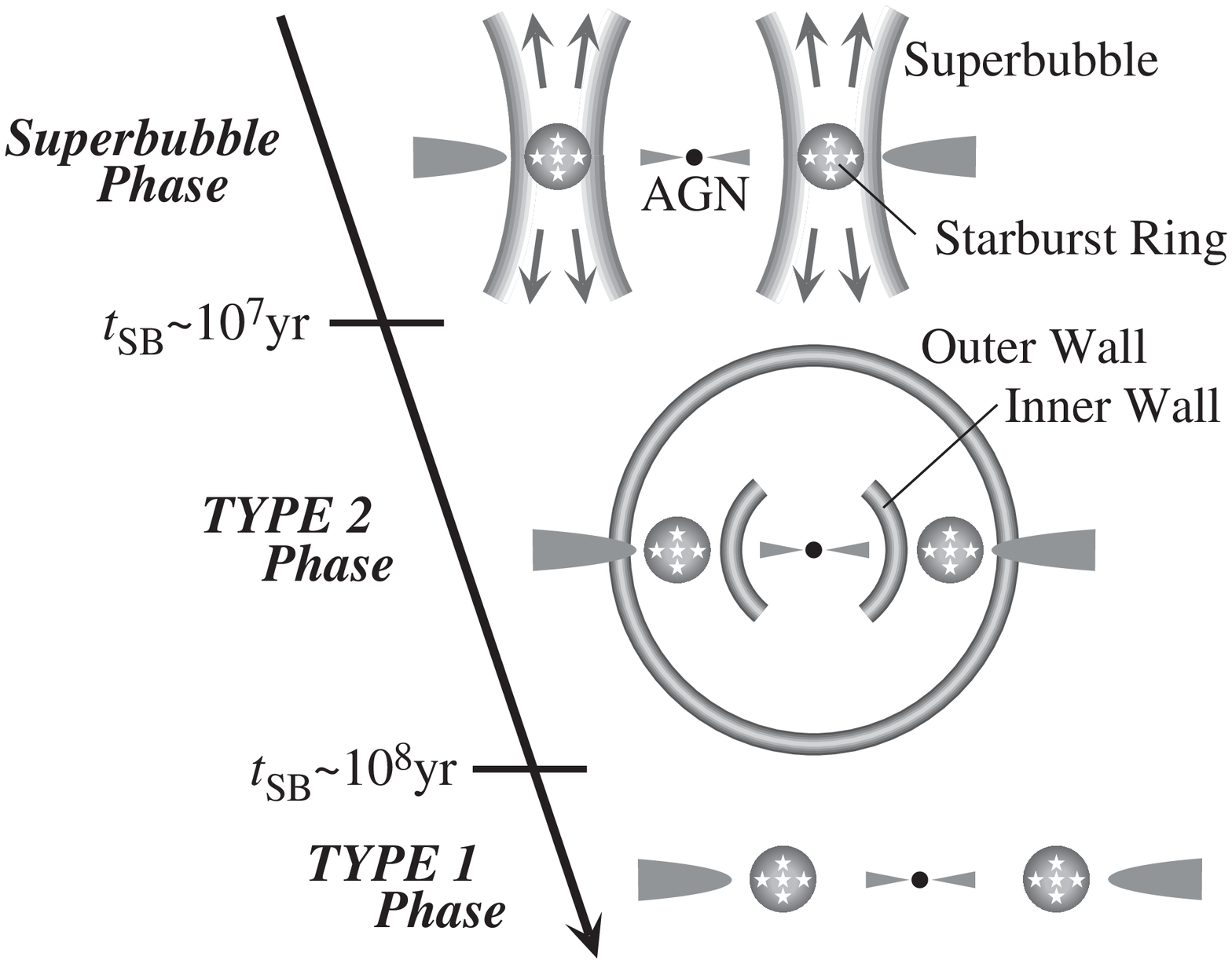,width=3.4in}}
\figcaption[fig7.eps] {
Schematic illustration of the evolution of circumnuclear structure.
Within $\sim 10^7$ yr, 
a large amount of dusty gas is blown away due to superbubbles
powered by type II supernovae in circumnuclear starburst regions.
The ejected dusty gas is supported by the radiation force by the AGN as well
as the starburst ring, and composes the obscuring walls.
Then, the nucleus is obscured by both walls and tends to be identified 
as type 2 AGN (Sy 2).
Based on the present stellar evolution model,
the nucleus tends to be observed as type 2 AGN (Sy 2) 
within several $\times 10^7$ yr,
since the possible $A_V$ decreases with time due to stellar evolution.
Finally, the walls disappear and 
the nucleus are observed as type 1 AGN (Sy 1).
Therefore, the AGN is destined to 
shift from type 2 to type 1 in several times $10^7$ yr in this picture.
However, 
QSOs are observed as type 1 AGN regardless of the age of the starburst, 
because the luminous AGN prevent the formation of the walls.
\label{fig7}}
\medskip
\noi
result, 
the dusty gas of $\gsim 10^8M_\odot$ would be supplied
within 
a typical lifetime of a superbubble ($\sim 10^7$ yr)
and composes the obscuring walls in the next stage.
However, an outer obscuring wall with $A_V \gsim$ 10
may not form, because it requires
the dusty gas of $\gsim 10^8 M_\odot$. 
On the other hand,
the required mass for an inner wall
is merely $\sim 10^6 M_\odot$.
The formation timescales of an inner wall and that of an outer wall
are several $\times 10^5$ yr and several $\times 10^6$ yr, respectively.
Both timescales are shorter than the evolutionary timescale of 
the starburst on the order of $\sim 10^7$ yr.
Thus, 
the nucleus is likely to be obscured by the inner wall with $A_V \gsim$ several
as well as the outer wall with $A_V \lsim$ a few,
whereas the possible $A_V$ decreases with time
due to the stellar evolution in the starburst regions.
Based on the present stellar evolution model,
the nucleus is heavily obscured and tends to be observed as type 2 AGN (Sy 2)
within several $\times 10^7$ yr.
Also, the inner wall 
would not be destroyed 
even if the superbubbles occur in this stage, 
since the bubble diameter on the equatorial plane is 
comparable to the scale height of the molecular disk
(Schiano 1985; Mac Low \& McCray 1988; 
Tomisaka \& Ikeuchi 1988; Mac Low, McCray, \& Norman 1989).

In the final stage, the obscuring walls disappear 
due to the diminution of the bolometric luminosity of the circumnuclear starburst.
Then, the nucleus would be identified as type 1 AGN (Sy 1).
To conclude, the AGN type is time-dependent and 
is destined to shift from type 2 to type 1 in several times $10^7$ yr.
It is stressed that 
no obscuring wall forms in the case of the luminous AGN like QSOs
so that type of QSO does not vary, and which is observed as type 1 
regardless of the age of the starburst.

Here we have assumed that AGN activity and the circumnuclear 
starbursts are simultaneous events.
From an observational point of view, it seems to be reasonable,
because it is revealed that 
AGN events are frequently accompanied by starbursts
in a variety of data
(Scoville et al. 1986;
Soifer et al. 1986;
Heckman et al. 1989, 1995;
Marconi et al. 1994; 
Neff et al. 1994;
Cid Fernandes \& Terlevich 1995;
Genzel et al. 1995;
Maiolino et al. 1995, 1998a;
Oliva et al. 1995;
Storchi-Bergmann et al. 1996;
Rodriguez-Espinosa, Rudy \& Jones 1987;
Elmouttie et al. 1998;
Cid Fernandes, Storchi-Bergmann, \& Schmitt 1998).
In addition, as a physical solution which links the two events,
Norman \& Scoville (1988)
have suggested a model,
whereby the interstellar medium
which is released by post-main-sequence stars in starburst regions
feeds a central black hole.
Also, the radiatively driven mass accretion onto a black hole
due to the radiation drag,
which is proposed by 
Umemura, Fukue, \& Mineshige (1997, 1998),
Fukue, Umemura, \& Mineshige (1997),
and Ohsuga et al. (1999),
would link the two events.

In the present analysis, we have assumed the interstellar gas containing 
a large amount of dust which is inferred in QSOs
(Barvainis, Antonucci, \& Coleman 1992; Ohta et al. 1996; Omont et al. 1996;
Dietrich \& Wilhelm-Erkens 2000).
However, the dust grains could be destroyed in shock-heated hot
gas and they could be reproduced in cooled superbubble winds.
The dust properties around AGNs are under debate also from an observational
point of view.
The destruction, formation, and growth of the dust grains 
would make obviously significant effects not only 
in the present picture but also in a generic problem of obscuration.
Thus, this issue should be considered more carefully 
in the future analysis.

Finally, we discuss the stabilities of the obscuring walls.
First of all, the walls are unlikely to be subject to the Rayleigh-Taylor 
instability. As shown in Figure 1,
the density gradient of the inner wall is positive inside the density peak
and negative outside the peak.
As shown in Figure 3, the acceleration works in the same directions
as the density gradient in both sides.
Thus, the Rayleigh-Taylor instability does not occur 
at the inner wall.
In the case of the outer wall,
the radiation force by the AGN and the starburst ring,
which pushes up the wall, is the dominant force at the inner surface,
but the dusty gas is attracted by the gravity at the outer surface.
As a result, both walls turn out to be stable to the Rayleigh-Taylor mode.
However, the walls might be subject to the other instabilities.
When the density perturbations occur in the walls, 
the thick parts of the walls go down by the gravity, whereas 
the radiation force lifts the less thick parts,
since the radiation force is ineffective for heavily optically-thick parts.
Then, the walls are likely to have more complicated configurations.
Moreover, thermal instabilities might break out,
because the radiative cooling works at the dense regions effectively,
whereas the gas remains as warm as $\sim 10^4$ K in the
optically thin regions due to radiative heating.
Simultaneously, the self-gravitating instability might grow at the cool 
and dense parts of the walls. 
Also, the actual circumnuclear starburst is not
perfectly ring-shaped but more or less clumpy and therefore makes
non-axisymmetric radiation fields, although the wheel effects by
the rotation could smear out non-axisymmetric effects to some degree 
(see also Ohsuga \& Umemura 1999).
In addition, the dusty walls are possibly damaged due to the collision of
starburst winds or dusty gas expelled from mass losing stars inside the walls.
Consequently,
the walls could be distorted and fragment into small pieces.
How the gas clouds resulting from the fragmentation
obscure the AGN is an interesting issue,
but the details are not clear before radiation-hydrodynamical simulations
are performed.

\section{CONCLUSIONS}

In the present analysis,
by taking optical depth due to dust opacity into consideration,
we have analyzed the stable equilibrium configuration of dusty gas
in the circumnuclear regions 
in radiation fields by a starburst and an AGN.
It has been found that 
the radiation pressure by a circumnuclear starburst sustains
an inner obscuring wall of several ten parsecs 
and an outer obscuring wall of several hundred parsecs.
The total extinction of the obscuring walls has turned out to be around ten.
Then, the nucleus would be identified as type 2.
If a starburst is intrinsically faint or becomes dimmer owing to
the stellar evolution there, neither an outer wall nor inner one
would not form. Then, the nucleus could be observed as a type
1 AGN (Sy 1). 
The results on outer and inner walls share a certain similarity
in that the strong radiation force by the AGN prevents the formation of
an outer wall as well as an inner wall.
These results provide a physical explanation
for the correlation between AGN type and host properties, whereby
Seyfert type 2 galaxies 
are more frequently associated with circumnuclear starbursts 
than type 1 galaxies, whereas QSOs are mostly observed as type 1 
even though vigorous star-forming activities are often observed
in the host galaxies.
%
Also, the large-scale outer wall of several hundred parsecs
is consistent with the observations on
obscuring material extending up to $\geq 100$ pc around the nuclei.
%
%
Finally, 
this model predicts the time evolution of the AGN type from type 2 to type 1
in several times $10^7$ yr
as a circumnuclear starburst becomes dimmer due to stellar evolution.

\acknowledgments

We are grateful to T. Nakamoto, and H. Susa, for helpful discussion.
We thank also the anonymous referee for valuable comments.
The calculations were carried out at Center for Computational Physics 
in University of Tsukuba. This work is 
supported in part by Research Fellowships of the Japan Society
for the Promotion of Science for Young Scientists, 6957 (KO)
and the Grants-in Aid of the
Ministry of Education, Science, Culture, and Sport, 09874055 (MU).

\end{document}